\documentclass[aps,prd,twocolumn,10pt,superscriptaddress,nofootinbib,nobibnotes,longbibliography]{revtex4-1}

\usepackage{amssymb}
\usepackage{graphicx}
\usepackage{amsmath}
\usepackage{hyperref}
\usepackage{subfigure}
\usepackage{multirow}
\usepackage{setspace}
\usepackage{verbatim}
\usepackage{float}
\usepackage{color}
\usepackage{ulem}
\usepackage[utf8]{inputenc}
\usepackage[table,xcdraw]{xcolor}
\usepackage{makecell}
\usepackage{url}

\begin{document}

\title{No-go guide for the Hubble tension: late-time or local-scale new physics}

\author{Lu Huang}
\email{huanglu@itp.ac.cn}
\affiliation{CAS Key Laboratory of Theoretical Physics, Institute of Theoretical Physics, Chinese Academy of Sciences (CAS), Beijing 100190, China}

\author{Shao-Jiang Wang}
\email{schwang@itp.ac.cn (corresponding author)}
\affiliation{CAS Key Laboratory of Theoretical Physics, Institute of Theoretical Physics, Chinese Academy of Sciences (CAS), Beijing 100190, China}
\affiliation{Asia Pacific Center for Theoretical Physics (APCTP), Pohang 37673, Korea}

\author{Wang-Wei Yu}
\email{wangwei.yu@aei.mpg.de}
\affiliation{CAS Key Laboratory of Theoretical Physics, Institute of Theoretical Physics, Chinese Academy of Sciences (CAS), Beijing 100190, China}
\affiliation{Max-Planck-Institut f{\"u}r Gravitationsphysik (Albert-Einstein-Institut), Callinstra{\ss}e 38, D-30167 Hannover, Germany}
\affiliation{Leibniz Universit{\"a}t Hannover, 30167 Hannover, Germany}


\begin{abstract}
The standard model of modern cosmology might be cracked by the recent persistent hot debate on the Hubble-constant ($H_0$) tension, which manifests itself as the sound-horizon ($r_s$) tension or absolute-magnitude ($M_B$) tension if deeming the origin of the Hubble tension from modifying the early or late Universe, respectively. In this paper, we achieve a fully model-independent constraint (fitting a model-independent global parameterization to a model-independent inverse distant ladder with a model-independent high-redshift calibration) on late-time models with strong evidence against homogeneous new physics over the $\Lambda$-cold-dark-matter ($\Lambda$CDM) model. Further using this model-independent constraint to calibrate sufficiently local supernovae with corresponding late-time models extrapolated below the homogeneity scale, we find surprisingly that, although both $H_0$ tension and $M_B$ tension are absent in our local Universe, a combination of $H_0$ and $M_B$ as the intercept $a_B$ of the magnitude-redshift relation exhibits $3\sim 7\sigma$ tension even for the $\Lambda$CDM model. This $a_B$ tension seems to call for local-scale inhomogeneous new physics disguised as local observational systematics.
\end{abstract}
\maketitle


\section{Introduction}

The $\Lambda$-cold-dark-matter ($\Lambda$CDM) model as the standard model of modern cosmology has been well established as a phenomenologically concordant model~\cite{Planck:2013pxb}, fitting simultaneously into the current precision cosmology~\cite{Moresco:2022phi} including cosmic microwave background (CMB), baryon acoustic oscillation (BAO), and type Ia supernovae (SNe Ia). This concordant model might be cracked by an ever-enlarging tension~\cite{Riess:2016jrr,Riess:2018byc,Riess:2018uxu,Riess:2019cxk,Riess:2020fzl} on the Hubble constant $H_0$ between the global fitting of Planck 2018 observation with $\Lambda$CDM extrapolation~\cite{Planck:2018vyg} and the local quasi-direct measurements from Pantheon+ compilation~\cite{Scolnic:2021amr} with SH0ES (Supernovae and $H_0$ for the Equation of State of dark energy) calibration~\cite{Riess:2021jrx}. If not caused by any of the early-time or late-time systematics, the above Hubble-constant tension~\cite{Bernal:2016gxb,Verde:2019ivm,Knox:2019rjx,Riess:2020sih,Freedman:2021ahq} could be a promising clue to the new physics. However, despite numerous attempts claiming to relieve or even resolve this \textit{$H_0$ tension}~\cite{DiValentino:2020zio,DiValentino:2021izs,Schoneberg:2021qvd,Shah:2021onj,Abdalla:2022yfr}, various no-go arguments have been suggested against $\Lambda$CDM extensions~\cite{Guo:2018ans} and new physics from either the early Universe~\cite{Krishnan:2020obg,Jedamzik:2020zmd,Hill:2020osr,Lin:2021sfs,Vagnozzi:2021gjh,Philcox:2022sgj} or the late Universe~\cite{Benevento:2020fev,Camarena:2021jlr,Efstathiou:2021ocp,Cai:2021weh,Cai:2022dkh}. On the other hand, if the $\Lambda$CDM model should indeed be modified, the Hubble tension cannot be the only tension that manifests the breakdown of the current standard model of cosmology. Below we only mention the most relevant ones to the $H_0$ tension but refer most of the other tensions to the most recent update~\cite{Perivolaropoulos:2021jda}.

\textbf{\textit{$r_s$ tension.---}}
One such tension comes from tracing the origin of the Hubble tension back to the early Universe but assuming the $\Lambda$CDM model at the late time. After extrapolated into higher redshifts with the aid of BAO data~\cite{BOSS:2016wmc,eBOSS:2020yzd,DES:2021esc}, the SNe Ia sample calibrated by SH0ES measurements usually infers a smaller sound horizon $r_s$ than that constrained by the CMB data alone~\cite{Vonlanthen:2010cd,Audren:2012wb,Verde:2016wmz,Aylor:2018drw}. As BAO data apparently maintains a nearly constant product $r_sH_0$, this \textit{$r_s$ tension}~\cite{Bernal:2016gxb,Verde:2019ivm,Knox:2019rjx,Riess:2020sih} also serves as an early-Universe reflection of the $H_0$ tension, whose resolutions~\cite{Khalife:2023qbu} naively require a reduced sound horizon from modifying either early expansion or recombination histories~\cite{Jedamzik:2020zmd}. However, such early resolutions necessarily either suppress or shorten the growth rate or growing time for matter perturbations, which can only be compensated with a larger matter fraction to match the late-time galaxy clustering or lensing data~\cite{Jedamzik:2020zmd,Hill:2020osr} but yet commonly worsen the dubbed \textit{$S_8$ tension}~\cite{DiValentino:2020vvd,Perivolaropoulos:2021jda,Abdalla:2022yfr}, calling for additional decaying matter mechanisms to rescue at the late time~\cite{Vattis:2019efj,Pandey:2019plg,Clark:2020miy}. Besides, these early resolutions also intimately tend to prefer an extremely scale-invariant scalar spectrum index $n_s\approx1$~\cite{Poulin:2018cxd,Niedermann:2019olb,Niedermann:2020dwg,Ye:2020btb,Ye:2021nej,Jiang:2022uyg,Jiang:2022qlj,Peng:2023bik}, calling for even more exotic inflationary model buildings~\cite{Braglia:2020bym,Takahashi:2021bti,Ye:2022efx,Braglia:2022phb,Lin:2022gbl,Fu:2023tfo}. Therefore, if the Hubble tension indeed originates from the early Universe, then it is never enough just to alter the early Universe alone~\cite{Vagnozzi:2023nrq}, but to specifically fine-tune both the late and primordial Universe at the same time, which is way too much price to pay to rebuild the concordant cosmology.

\textbf{\textit{$M_B$ tension.---}}
The other tension comes from locating the cause of the Hubble tension in the late Universe but assuming the $\Lambda$CDM model at the early time. After calibrated by the sound horizon from CMB constraints, the inverse distance ladder (IDL)~\cite{Cuesta:2014asa,Heavens:2014rja,Aubourg:2014yra,Verde:2016ccp,Alam:2016hwk,Verde:2016wmz,Macaulay:2018fxi,Feeney:2018mkj,Lemos:2018smw,eBOSS:2020yzd} consisting of SNe Ia and BAO data usually leads to a fainter absolute magnitude $M_B$ than SH0ES measurements for the SNe Ia. As $M_B$ and $H_0$ jointly form the intercept $a_B$ of the magnitude-redshift relation constrained by SNe Ia data, this \textit{$M_B$ tension} also serves as a late-Universe reflection of the Hubble tension, whose resolutions seem to require either a fainter absolute magnitude from local-Universe systematics or a faster expansion rate of late Universe, for example, a phantom-like dark energy (PDE) transition~\cite{Mortonson:2009qq}. However, such a late-Universe PDE transition is strongly constrained by the current IDL data calibrated by the CMB prior to the sound horizon~\cite{Benevento:2020fev,Camarena:2021jlr,Efstathiou:2021ocp}. Moreover, other late-Universe deviations from the $\Lambda$CDM model are also strongly disfavored by an improved IDL method~\cite{Cai:2021weh,Cai:2022dkh}, where the model-independent fitting model adopts a global, economic, and precise Parameterization based on the cosmic Age (PAge)~\cite{Huang:2020mub,Luo:2020ufj,Huang:2020evj,Huang:2021aku,Huang:2021tvo,Huang:2022txw,Li:2022inq,Wang:2023mir,Wang:2024nsi,Wang:2024rus} while high-redshift calibrators employ cosmologically model-independent Hubble parameters from cosmic chronometer (CC) data~\cite{Jimenez:2001gg}.

The above-mentioned $r_s$ tension, the associated $S_8$ tension, and the $M_B$ tension have narrowed down a satisfactory resolution to the local Universe from either local systematic errors or local new physics. The local systematics have been thoroughly investigated by the SH0ES team~\cite{Riess:2021jrx}, finding no significant contribution to the current $M_B$ tension. Therefore, it seems to further pin down the local-Universe resolutions to local new physics. In this paper, we will first further strengthen the so-called~\cite{Vagnozzi:2023nrq} late-time ``No-Go theorem''~\cite{Benevento:2020fev,Camarena:2021jlr,Efstathiou:2021ocp,Cai:2021weh,Cai:2022dkh} by using not only the model-independent PAge model and model-independent CC calibration but also a fully model-independent IDL dataset consisting of Hubble-flow (HF) SNe Ia and two-dimensional (2D) BAO data, and then use this fully model-independent IDL constraints to calibrate the local SNe sample, revealing a new tension in the intercept $a_B$ of magnitude-redshift relation between the late Universe and local Universe.

\section{Model-independent IDL dataset}

A local distance ladder is used for calibrating a distant sample of distance indicators (e.g. SNe Ia), while the ``inverse'' distance ladder refers to a reverse process by calibrating a combined distant sample of distance indicators (i.e. SNe+BAO) with more distant objects, for example, the sound horizon $r_s$ from CMB~\cite{Vonlanthen:2010cd,Audren:2012wb,Verde:2016wmz,Aylor:2018drw} or time-delay distance $D_{\Delta t}$~\cite{Arendse:2019hev} from strong lensing time delay~\cite{Wong:2019kwg}, which still separately depends on early or late/local Universe via integrating over the expansion history in $r_s$ or $D_{\Delta t}$. We propose here a fully model-independent IDL method by adopting following fully model-independent datasets:

\subsection{HF SNe Ia data}

The HF SNe Ia are usually selected in the redshift range $0.0233<z<0.15$, where the lower bound $z_{\min}=0.0233$ corresponds to the homogeneity scale $R_\mathrm{homo}\approx70\,\mathrm{Mpc}/h$~\cite{Scrimgeour:2012wt} since it is required to suppress the cosmic variance~\cite{1992AJ....103.1427T,Shi:1995nq,Shi:1997aa,Wang:1997tp} from our local matter density contrast~\cite{Kenworthy:2019qwq} within $R_\mathrm{homo}$~\cite{Sinclair:2010sb,Marra:2013rba,Ben-Dayan:2014swa,Camarena:2018nbr}, while the upper bound $z_{\max}=0.15$ is required to detach any cosmological dependence on the late-Universe evolution. Specifically, we select 490 HF SNe Ia out of a state-of-art sample from the Pantheon+ compilation~\cite{Riess:2021jrx,Brout:2022vxf,Scolnic:2021amr,Brout:2021mpj,Peterson:2021hel,Carr:2021lcj,Popovic:2021yuo}, which mainly adds low-redshift ($z<0.1$) SNe from extra low-$z$ surveys (see, e.g.~\cite{Scolnic:2021amr}) compared to the original Pantheon sample~\cite{Scolnic:2017caz}. The full data releases of Pantheon+ sample are publicly available at \href{https://pantheonplussh0es.github.io/}{https://pantheonplussh0es.github.io/}. The likelihood $\ln\mathcal{L}_\mathrm{HFSN}=-\frac12\mathbf{\Delta}_\mathrm{HFSN}^\mathrm{T}C_\mathrm{stat.+syst.}^{-1}\mathbf{\Delta}_\mathrm{HFSN}$ is constructed from the distance-modulus residual vector $\Delta_{\mathrm{HFSN},i}=(m_{B,i}-M_B)-\mu_\mathrm{model}(z_i)$, where the apparent B-band peak magnitude $m_{B,i}$ of $i$-th SN at redshift $z_i$ has been corrected for the stretch, color, simulation bias, and mass-step effects, and the absolute B-band peak magnitude $M_B$ will be left free or calibrated by either the SH0ES measurement or the IDL constraint later, while the distance modulus $\mu_\mathrm{model}(z_i)$ at corresponding redshift is modeled as
\begin{align}\label{eq:distance-moduli}
\mu_{\mathrm{model}}(z_i) =5\lg d_L(z_i)+5\lg \frac{c}{H_0 \cdot\mathrm{Mpc}}+25, 
\end{align}
where the dimensionless luminosity distance $d_L(z)\equiv D_L(z)/(c/H_0)=(1+z_\mathrm{hel})\int_{0}^{z_\mathrm{HD}}\mathrm{d}z'/E(z')$ is computed given a cosmological model for the dimensionless Hubble parameter $E(z)\equiv H(z)/H_0$. Here $z_\mathrm{hel}$ is the heliocentric redshift and $z_{\mathrm{HD}}$ is the Hubble-diagram redshift with CMB and peculiar velocity corrections in Pantheon+ samples. The covariance matrix $C_{\mathrm{stat.+syst.}}$ includes all statistical and systematic uncertainties as well as the expected light-curve correlations in the sample.

\subsection{2D BAO dataset}

The BAO from the early Universe leaves permanent imprints on the galaxy clustering in the late Universe, extremalizing locally the two-point correlation function (2PCF) of galaxy pairs with a comoving separation of sound-horizon size $r_d$ at drag epoch. Due to the redshift-space distortion~\cite{Kaiser:1987qv} and Alcock-Paczynski effect~\cite{Alcock:1979mp} for a mismatched fiducial cosmology, the usual anisotropic fitting to the measured BAO signals with templates~\cite{Xu:2012fw} for the monopole and quadrupole moments of 2PCF can provide independent measurements~\cite{Okumura:2007br,Gaztanaga:2008xz,Blake:2011ep,Chuang:2011fy} for the Hubble parameter $H(z)$ and angular diameter distance $D_{A}(z)\equiv D_L(z)/(1+z)^2$ with respect to the fiducial cosmology. Although this kind of relative measurements can largely eliminate the explicit dependence on the fiducial cosmology, the model-dependent bias can still sneak into analysis in following three aspects: (i) differences in the distance-redshift relation used for calculating the galactic separation between the true and fiducial cosmologies~\cite{Heinesen:2019phg} can scale non-linearly in some inhomogeneous models or with backreaction; (ii) differences in modeling the comoving clustering between the true and fiducial cosmologies~\cite{Carter:2019ulk} might not be fully absorbed into the same set of nuisance parameters that are used to marginalize over any non-BAO signal~\cite{Wang:2017mcf} in the fitting template; (iii) differences in estimating the covariance matrix from mock random samples~\cite{Wang:2017mcf} can be quite different from numerical simulations of large scale structures beyond the standard cosmological model~\cite{Bolejko:2017fos}.

\begin{table}[H]
\caption{2D BAO data}
\label{tab:2DBAO}
\renewcommand{\arraystretch}{1.2}
\centering
\begin{tabular}{|c|c|c|r|}
\hline
\hline
$z$ & $\theta_{\mathrm{BAO}}[\mathrm{deg}]$ & $\sigma_{\mathrm{BAO}}[\mathrm{deg}]$ & References \\
\hline
\hline
$0.11$ & $19.8$ & $1.05$ &~\cite{deCarvalho:2021azj}\\
$0.235$ & $9.06$ & $0.23$ &~\cite{Alcaniz:2016ryy}\\
$0.365$ & $6.33$ & $0.22$ &~\cite{Alcaniz:2016ryy}\\
$0.45$ & $4.77$ & $0.17$ &~\cite{Carvalho:2015ica}\\
$0.47$ & $5.02$ & $0.25$ &~\cite{Carvalho:2015ica}\\
$0.49$ & $4.99$ & $0.21$ &~\cite{Carvalho:2015ica}\\
$0.51$ & $4.81$ & $0.17$ &~\cite{Carvalho:2015ica}\\
$0.53$ & $4.29$ & $0.30$ &~\cite{Carvalho:2015ica}\\
$0.55$ & $4.25$ & $0.25$ &~\cite{Carvalho:2015ica}\\
$0.57$ & $4.59$ & $0.36$ &~\cite{Carvalho:2017tuu}\\
$0.59$ & $4.39$ & $0.33$ &~\cite{Carvalho:2017tuu}\\
$0.61$ & $3.85$ & $0.31$ &~\cite{Carvalho:2017tuu}\\
$0.63$ & $3.90$ & $0.43$ &~\cite{Carvalho:2017tuu}\\
$0.65$ & $3.55$ & $0.16$ &~\cite{Carvalho:2017tuu}\\
$2.225$ & $1.82$ & $0.21$ &~\cite{deCarvalho:2020ftb}\\
\hline
\hline
\end{tabular}
\end{table}

However, unlike the conventional BAO that explicitly depends on the fiducial cosmology, the 2D BAO~\cite{Carvalho:2015ica} from measuring the two-point angular correlation function relies only on the angular separation between pairs of galaxies without assuming a fiducial cosmology, rendering the most model-independent constraint on angular BAO scales $\theta_\mathrm{BAO}(z_i)$ modeled as
\begin{align}\label{eq:angular-scale-measurement}
 \theta_\mathrm{model}(z_i)= \frac{r_{d}}{(1+z_i)D_{A}(z_i)}, 
\end{align}
which are totally uncorrelated between any two different redshifts $z_i$ since they are obtained from totally disjoint spherical shells with the redshift thickness $\Delta z$ just enough to extract significant transversal BAO signals, and hence the covariance matrix includes only diagonal errors $\sigma_\mathrm{BAO}(z_i)$~\cite{Anselmi:2018vjz}. Therefore, the likelihood is estimated as $\ln\mathcal{L}_\mathrm{2DBAO}=-\frac12\sum_i(\theta_\mathrm{BAO}(z_i)-\theta_\mathrm{model}(z_i))^2/\sigma_\mathrm{BAO}(z_i)^2$ from 15 data points of 2D BAO as shown in Tab.~\ref{tab:2DBAO} collected from Refs.~\cite{deCarvalho:2021azj,Alcaniz:2016ryy,Carvalho:2015ica,Carvalho:2017tuu,deCarvalho:2020ftb} first applied to Refs.~\cite{Nunes:2020hzy,Nunes:2020uex}. See also, e.g., Refs.~\cite{Bernui:2023byc,Lemos:2023qoy,Gomez-Valent:2024tdb,Favale:2024sdq,Ruchika:2024lgi} for recent studies using 2D BAO data.

\subsection{CC calibration}

The CC data as high-redshift calibrations to inversely calibrate the late-Universe distance ladder HFSN+2DBAO is completely independent of both early and late/local Universe as the Hubble parameter $H(z)$ is directly measured by~\cite{Jimenez:2001gg}
\begin{align}
H(z)=-\frac{1}{1+z}\frac{\mathrm{d}z}{\mathrm{d}t}=-\frac{1}{1+z_\mathrm{eff}}\frac{\Delta z}{\Delta t}
\end{align}
for passively-evolving early-time massive galaxies that formed around the same time with much shorter age differences $\Delta t$ than their evolving time scales, and are separated by a small redshift interval $\Delta z$  around $z_\mathrm{eff}$. The likelihood $\ln\mathcal{L}_\mathrm{CC}=-\frac12\mathbf{\Delta}_\mathrm{CC}^\mathrm{T}C_\mathrm{stat.+syst.}^{-1}\mathbf{\Delta}_\mathrm{CC}$ is constructed from the residual vector $\Delta_{\mathrm{CC},i}=H_\mathrm{obs}(z_i)-H_\mathrm{model}(z_i)$ and a completed covariance matrix~\cite{Moresco:2020fbm,Moresco:2022phi} including both statistical/systematic uncertainties, where the observed Hubble parameters $H_\mathrm{obs}(z_i)$ as shown in Tab.~\ref{tab:CC} are adopted from Refs.~\cite{Jimenez:2003iv,Simon:2004tf,Stern:2009ep,Moresco:2012jh,Zhang:2012mp,Moresco:2015cya,Moresco:2016mzx,Ratsimbazafy:2017vga,Borghi:2021rft}, while the full covariance matrix $C_\mathrm{stat.+syst.}$ can be found  at ~\href{https://gitlab.com/mmoresco/CCcovariance}{https://gitlab.com/mmoresco/CCcovariance}.

\begin{table}[H]
\caption{CC data}
\label{tab:CC}
\renewcommand{\arraystretch}{1.2}
\centering
\begin{tabular}{|c|c|c|}
\hline
\hline
$z$ & $H(z)$ km/s/Mpc  & References \\
\hline
\hline
0.1 & $69 \pm 12$ &~\cite{Jimenez:2003iv,Stern:2009ep}  \\
0.17 & $83 \pm 8$ &~\cite{Simon:2004tf,Stern:2009ep} \\
0.27 & $77 \pm 14$ &~\cite{Simon:2004tf,Stern:2009ep} \\
0.4 & $95 \pm 17$ &~\cite{Simon:2004tf,Stern:2009ep}  \\
0.48 & $97 \pm 62$ &~\cite{Stern:2009ep}  \\
0.88 & $90 \pm 40$ &~\cite{Stern:2009ep} \\
0.9 & $117 \pm 23$ &~\cite{Simon:2004tf,Stern:2009ep}  \\
1.3 & $168 \pm 17$ &~\cite{Simon:2004tf,Stern:2009ep} \\
1.43 & $177 \pm 18$ &~\cite{Simon:2004tf,Stern:2009ep} \\
1.53 & $140 \pm 14$ &~\cite{Simon:2004tf,Stern:2009ep} \\
1.75 & $202 \pm 40$ &~\cite{Simon:2004tf,Stern:2009ep} \\
\hline
\hline
0.1791 & $75 \pm 4$ &~\cite{Moresco:2012jh} \\
0.1993 & $75 \pm 5$ &~\cite{Moresco:2012jh} \\
0.3519 & $83 \pm 14$ &~\cite{Moresco:2012jh} \\
0.5929 & $104 \pm 13$ &~\cite{Moresco:2012jh} \\
0.6797 & $92 \pm 8$ &~\cite{Moresco:2012jh} \\
0.7812 & $105 \pm 12$ &~\cite{Moresco:2012jh} \\
0.8754 & $125 \pm 17$ &~\cite{Moresco:2012jh} \\
1.037 & $154 \pm 20$ &~\cite{Moresco:2012jh} \\
\hline
\hline
0.07 & $69.0 \pm 19.6$ &~\cite{Zhang:2012mp} \\
0.12 & $68.6 \pm 26.2$ &~\cite{Zhang:2012mp} \\
0.20 & $72.9 \pm 29.6$ &~\cite{Zhang:2012mp} \\
0.28 & $88.8 \pm 36.6$ &~\cite{Zhang:2012mp} \\
\hline
\hline
1.363 & $160 \pm 33.6$ &~\cite{Moresco:2015cya} \\
1.965 & $186.5 \pm 50.4$ &~\cite{Moresco:2015cya} \\
\hline
\hline
0.3802 & $83.0 \pm 13.5$ &~\cite{Moresco:2016mzx} \\
0.4004 & $77.0 \pm 10.2$ &~\cite{Moresco:2016mzx} \\
0.4247 & $87.1 \pm 11.2$ &~\cite{Moresco:2016mzx} \\
0.4497 & $92.8 \pm 12.9$ &~\cite{Moresco:2016mzx} \\
0.4783 & $80.9 \pm 9$ &~\cite{Moresco:2016mzx} \\
\hline
\hline
0.47 & $89 \pm 49.6$ &~\cite{Ratsimbazafy:2017vga}\\
\hline
\hline
0.75 & $98.8 \pm 33.6$ &~\cite{Borghi:2021rft} \\
\hline
\hline
\end{tabular}
\end{table}

\section{Model-independent parameterization}

Another indispensable pillar of the fully model-independent IDL method is to fit with a model-independent parameterization. As shown in Refs.~\cite{Cai:2021weh,Cai:2022dkh}, the traditional Taylor expansion in redshift $z$ or $y$-redshift $y\equiv1-a=z/(1+z)$ even up to the fifth orders cannot faithfully approximate the model it claims to parameterize at redshift $z\gtrsim1$ even for the model as simple as the $\Lambda$CDM model. Hence, we adopt a better parameterization dubbed PAge model~\cite{Huang:2020mub} as a collection of relatively gentle modifications to the $\Lambda$CDM model with $E(z)^2\equiv H(z)^2/H_0^2=\Omega_m(1+z)^3+1-\Omega_m$, while using a specific PDE transition model as a typical example of more violent saltation in the late and local Universe. We sketch these two models below:

\subsection{PAge model}

The spirit of PAge model is to parameterize the accumulated quantity (like the cosmic age) instead of the instant quantity (like the expansion rate) as the latter one can change more dramatically and hence more difficult to parameterize than the former one. For a pure matter Universe, it holds $Ht-2/3=0=H_0t_0-2/3$ for the cosmic age $t$ and its current value $t_0$. Since the early radiation duration is negligible and the matter duration contributes most of the cosmic age, it is natural to parameterize various late-time homogeneous dark energy models as deviations from the pure matter Universe by Taylor expansions in the cosmic age as~\cite{Huang:2020mub,Luo:2020ufj}
\begin{align}
Ht-\frac23=\left(H_0t_0-\frac23(1+\eta)\right)\left(\frac{t}{t_0}\right)+\frac23\eta\left(\frac{t}{t_0}\right)^2,
\end{align}
where a free parameter $\eta$ is introduced to characterize such a deviation from a pure matter Universe with $\eta=0$, while $H_0t_0\equiv p_\mathrm{age}$ is another free parameter in the PAge model together with $\eta$ to solve for $E(z)\equiv H(z)/H_0$ with the help of the definition $H(z)\equiv-\mathrm{d}z/\mathrm{d}t/(1+z)$. Higher-order expansions like the cubic term in cosmic age~\cite{Huang:2021aku} are not necessary as shown in Ref.~\cite{Cai:2022dkh} as the quadratic order is precise enough to parameterize the usual late-time homogeneous models~\cite{Huang:2020mub,Luo:2020ufj} even up to a higher redshift $z\gtrsim1$~\cite{Cai:2021weh,Cai:2022dkh} where some of our CC calibrators live. For a given model to be parameterized by the PAge model, one is free to choose a moment when these two models coincide, which is usually convenient to be fixed at our present day when the two PAge parameters can be related by $\eta=1-\frac32p_\mathrm{age}^2(1+q_0)$ via the current value $q_0$ of the deceleration parameter $q(t)\equiv-\ddot{a}a/\dot{a}^2$. Therefore, the PAge model serves as a global, economic, and precise parameterization in the late/local Universe for any homogeneous modification gently beyond the $\Lambda$CDM model.

\subsection{PDE model}

For those more abrupt changes in the expansion rate that cannot be well captured by the PAge model, we can use the PDE model following Ref.~\cite{Efstathiou:2021ocp} as a representative illustration. The Hubble parameter reads
\begin{align}
H(z)=H_0^f\sqrt{\Omega_m(1+z)^3+(1-\Omega_m)\left(1+\Delta\,e^{-\left(\frac{z}{z_c}\right)^\beta}\right)},
\end{align}
where the phantom-like transition occurs at a transition reshift $z_c$ with a strength $\Delta$ and an index $\beta$, leading to a Hubble constant $H_0\equiv H_0^f\sqrt{\Omega_m+(1-\Omega_m)(1+\Delta)}$ generally larger than the fiducial one $H_0^f$ while still keeping its high-redshift behavior at $z\gg z_c$ as if it goes along with the $\Lambda$CDM model extrapolating at $H_0=H_0^f$.

\section{Late-Universe ``No-Go theorem''}

\begin{figure}[H]
\centering
\includegraphics[width=0.49\textwidth]{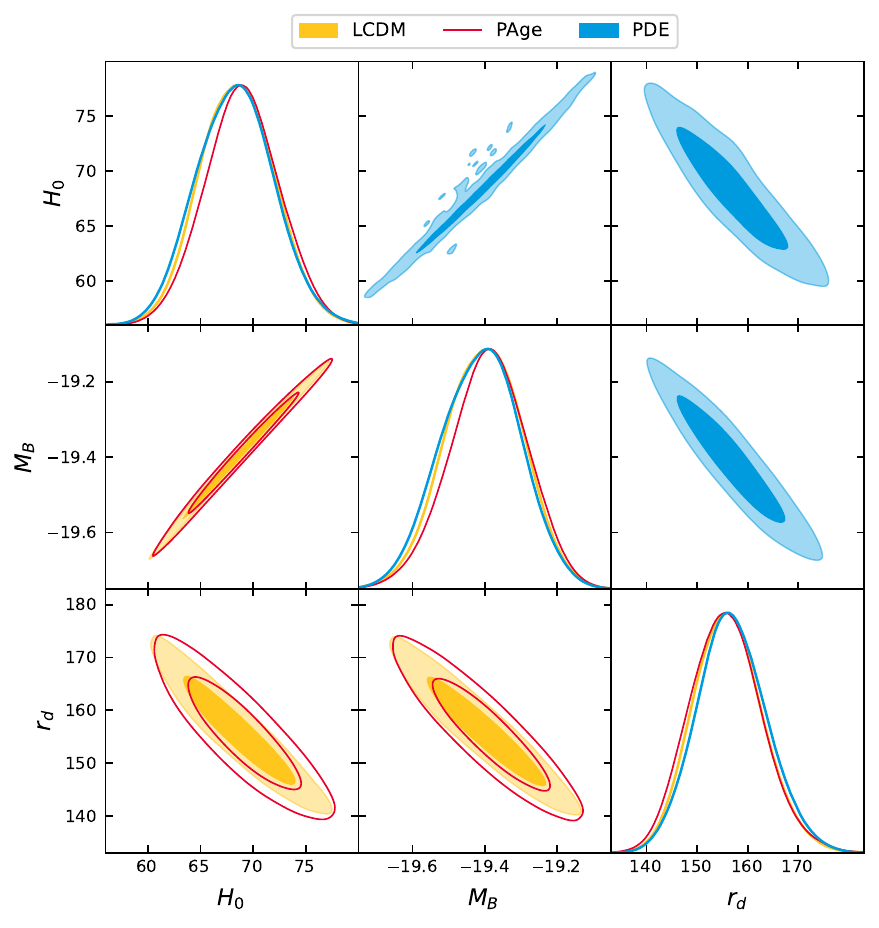}\\
\caption{The marginalized 1$\sigma$ and 2$\sigma$ constraints with the 1D posteriors on $H_0$, $M_B$, and $r_d$ from fitting $\Lambda$CDM, PAge and PDE models with the fully model-independent IDL dataset.}
\label{fig:H0-MB-rd}
\end{figure}

Fitting the fully model-independent IDL dataset from HF SN+2D BAO+CC to the $\Lambda$CDM, PAge, and PDE models in the late Universe ($z>0.0233$) with the Markov Chain Monte Carlo code \texttt{EMCEE}~\cite{Foreman-Mackey:2012any}, we can constrain their model parameters with flat priors presented in Tab.~\ref{tab:prior} from the joint likelihood $\ln\mathcal{L}_\mathrm{tot}=\ln\mathcal{L}_\mathrm{HFSN}+\ln\mathcal{L}_\mathrm{2DBAO}+\ln\mathcal{L}_\mathrm{CC}$. The full cosmological constraints are presented in Tab.~\ref{tab:FittingResultsLate}. In particular, the marginalized $1\sigma$ and $2\sigma$ constraints with 1D posteriors on parameters $H_0$, $M_B$, and $r_d$ are presented in Fig.~\ref{fig:H0-MB-rd} with almost indistinguishable differences for both PAge and PDE models with respect to the $\Lambda$CDM model.

\begin{table}[H]
\renewcommand{\arraystretch}{1.5}
\caption{Parameters and priors}
\label{tab:prior}
\centering
\begin{tabular}{|c|c|c|c|c|c|}
\hline
 \multicolumn{3}{|c|}{Model parameters} & & \multicolumn{2}{c|}{Parameter priors} \\
 \cline{1-3}\cline{5-6}
 $\Lambda$CDM & PAge & PDE & & Local Universe & Late Universe\\
 \hline
 \hline
 \multicolumn{3}{|c|}{$M_B$} & & \multicolumn{2}{c|}{$(-21, -18)$} \\
 \cline{1-3}\cline{5-6}
 \multicolumn{3}{|c|}{$r_d $} & & \multicolumn{2}{c|}{$(100, 200)$ [Mpc]} \\
 \cline{1-3}\cline{5-6}
 \multicolumn{2}{|c|}{$H_0$} & $H_0^f$ & & \multicolumn{2}{c|}{$(40, 90)$ [km/s/Mpc]} \\
 \cline{1-3}\cline{5-6}
 $\Omega_m$ & --- & $\Omega_m$ & & \multicolumn{2}{c|}{$(0, 1)$} \\
 \cline{1-3}\cline{5-6}
 --- & $p_{\rm{age}}$ & --- & & \multicolumn{2}{c|}{$(0.7, 1.2)$} \\
 \cline{1-3}\cline{5-6}
 --- & $\eta$ & --- & & \multicolumn{2}{c|}{$(-1, 1)$} \\
 \cline{1-3}\cline{5-6}
 --- & --- & $\Delta$ & & \multicolumn{2}{c|}{$(-1, 1)$} \\
 \cline{1-3}\cline{5-6}
 --- & --- & $z_c$ & & $(0, 0.0233)$ & $(0.0233, 1)$ \\
 \cline{1-3}\cline{5-6}
 --- & --- & $\beta$ & & \multicolumn{2}{c|}{$(-4, 4)$} \\
 \hline
\end{tabular}
\end{table}

This perfect match is unusual, as generally, the more parameters there are, the weaker the constraints on the parameters will be. Despite of similar constraints on the common parameters ($H_0$, $M_B$, and $r_d$), the PAge model has one more parameter than the $\Lambda$CDM model, effectively with $(p_\mathrm{age},\eta)$ in replacing of $\Omega_m$. In the PAge representation of the $\Lambda$CDM model, it holds $\eta=1-\frac32p_\mathrm{age}^2(1+q_0)$ if we match the PAge representation with the $\Lambda$CDM model at our present day, where the current deceleration parameter reads $q_0=-1+\frac32\Omega_m$ in the $\Lambda$CDM model. Then, it is easy to check that the effective parameter $\Omega_m=(1-\eta)/(\frac94p_\mathrm{age}^2)=0.31_{-0.11}^{+0.10}$ as a $\Lambda$CDM-equivalent analog from PAge+IDL posteriors is indeed weaker than the true $\Lambda$CDM+IDL constraint $\Omega_m=0.328_{-0.053}^{+0.044}$. As for the PDE model with much more parameters, the true Hubble constant $H_0\equiv H(z=0)$ is actually derived from all five model parameters $H_0^f$, $\Omega_m$, $\Delta$, $z_c$, and $\beta$. The direct constraint from PDE+IDL on $H_0^f=67.4\pm5.5$ km/s/Mpc is indeed weaker than the $\Lambda$CDM+IDL constraint $H_0=68.5\pm3.5$ km/s/Mpc, only the true Hubble constant $H_0=68.2\pm3.9$ km/s/Mpc has achieved conspiratorially a comparable constraint.

\begin{table}[H]
\caption{Constraints on late Universe from fitting IDL}
\label{tab:FittingResultsLate}
\renewcommand{\arraystretch}{1.5}
\centering
\begin{tabular}{|c|c|c|c|}
\hline
\multirow{2}{*}{Parameters} & \multicolumn{3}{c|}{$\text{Hubble flow SN Ia}+\text{2D BAO}+\text{CC}\equiv\text{IDL}$}        \\ \cline{2-4}  
& $\Lambda$CDM & PAge &PDE\\ \hline \hline

$H_0$ [km/s/Mpc] & $68.5\pm3.5$ & $69.0\pm 3.5$ & $68.2\pm3.9$\\ \hline
                  
$M_B$& $-19.40\pm0.11$ & $-19.39\pm0.11$ & $-19.41\pm0.11$ \\ \hline

$r_d$ [Mpc] & $156.1_{-7.3}^{+6.3}$ & $155.8\pm7.1$ & $156.9\pm6.9$ \\ \hline
                  
$\Omega_m$& $0.328_{-0.053}^{+0.044}$ & - & $0.348_{-0.088}^{+0.056}$ \\ \hline

$\Delta$& - & - & $0.09_{-0.43}^{+0.56}$ \\ \hline

$z_c$& - & - & $0.60_{-0.16}^{+0.37}$ \\ \hline

$\beta$& - & - & $-0.50_{-3.4}^{+4.2}$ \\ \hline

$p_{\rm age}$& - & $0.958_{-0.051}^{+0.041}$ & - \\ \hline

$\eta$& - & $0.36_{-0.20}^{+0.35}$ & - \\ \hline \hline

$\chi^2/\mathrm{d.o.f}$& $0.8669$ & $0.8685$ & $0.8710$ \\ \hline \hline

$\Delta \rm{BIC}$& - & $6.27$ & $18.42$ \\ \hline \hline

$\Delta \ln \mathcal{Z} $& - & $-0.553$ & $-0.707$ \\ \hline 
\end{tabular}
\end{table}

This model preference can be drawn more quantitatively from the Bayesian information criterion (BIC)~\cite{Schwarz:1978tpv} $\mathrm{BIC}=-2\ln\mathcal{L}_\mathrm{max}+k\ln N$ defined from a Gaussian approximation to the Bayesian evidence in the limit of large sample size, where the number of free parameters takes $k=4, 5, 7$ for $\Lambda$CDM, PAge, and PDE models, respectively, and the total number of data points $N=537$ consists of 490 HF SNe Ia, 15 2D BAO and 32 CC data points. We therefore find strong evidence ($\Delta\mathrm{BIC}=6.27$) against the PAge model and decisive evidence ($\Delta\mathrm{BIC}=18.42$) against the PDE model over the $\Lambda$CDM model. The Bayes factors~\cite{Trotta:2017wnx} against the PAge model ($\Delta\ln\mathcal{Z}=-0.553$) and the PDE model  ($\Delta\ln\mathcal{Z}=-0.707$) over the $\Lambda$CDM model are also substantial according to Jeffreys' scale~\cite{Jeffreys:1939xee} using \texttt{DYNESTY}~\cite{Speagle:2019ivv,Skilling:2004pqw}. This further strengthening the previous no-go arguments~\cite{Cai:2021weh,Cai:2022dkh} against any homogeneous new physics in the late Universe with the most model-independent manner so far. Replacing the 2D BAO with 3D BAO and the CC calibration with CMB-prior calibration would certainly improve the constraining power as done previously in earlier studies but weaken the late-time ``No-Go theorem'' with less model independence.

\section{Local-Universe $a_B$ tension}

Having deduced the late Universe cannot deviate even mildly away from the $\Lambda$CDM model, it is then tempting to further pin down the possible new physics within the local Universe ($z<0.0233$) by fitting to a local SN sample from 336 SNe Ia below the homogeneity scale in the Pantheon+ sample with well-separated peculiar-velocity corrections~\cite{Peterson:2021hel,Carr:2021lcj}. As it is the SH0ES calibration on $M_B$ that raises the issue of the Hubble tension in the first place, we will instead calibrate these local SNe Ia with the previously obtained fully model-independent IDL constraint on $M_B$. The full cosmological constraints are shown in Tab.~\ref{tab:FittingResultsLocal}. Note that in the $\Lambda$CDM model, the local SNe constrain $\Omega_m$ poorly, and hence we can also use the IDL $\Omega_m$ posterior as an extra prior in addition to the IDL $M_B$ posterior to calibrate the local SNe sample.

\begin{table}[H]
\caption{Constraints on local Universe with IDL $M_B$ prior}
\label{tab:FittingResultsLocal}
\renewcommand{\arraystretch}{1.5}
\centering
\begin{tabular}{|c|c|c|c|}
\hline
\multirow{2}{*}{Parameters} & \multicolumn{3}{c|}{IDL $M_B$ prior + Local SN Ia}        \\ \cline{2-4}  
& $\Lambda$CDM & PAge &PDE\\ \hline \hline

$H_0$ [km/s/Mpc] & $66.2\pm3.4$ & $66.6\pm 3.4$ & $50.0^{+7}_{-4}$\\ \hline
                  
$M_B$& $-19.40\pm0.11$ & $-19.39\pm0.11$ & $-19.41\pm0.11$ \\ \hline
                  
$\Omega_m$& $0.768_{-0.089}^{+0.23}$ & - & $0.27_{-0.26}^{+0.15}$ \\ \hline

$\Delta$& - & - & $0.70\pm0.16$ \\ \hline

$z_c$& - & - & $0.00218_{-0.00073}^{+0.00063}$ \\ \hline

$\beta$& - & - & $2.64_{-0.56}^{+1.3}$ \\ \hline

$p_{\rm age}$& - & $0.816_{-0.12}^{+0.019}$ & - \\ \hline

$\eta$& - & $-0.62_{-0.38}^{+0.14}$ & - \\ \hline \hline

$\chi^2/\mathrm{d.o.f}$& $1.460$ & $1.448$ & $1.308$ \\ \hline \hline

$\Delta \rm{BIC}$& - & $0.461$ & $-36.9$ \\ \hline \hline

$\Delta \ln  \mathcal{Z} $& - & $0.560$ & $16.4$ \\ \hline 
\end{tabular}
\end{table}

\begin{figure}[H]
\centering
\includegraphics[width=0.49\textwidth]{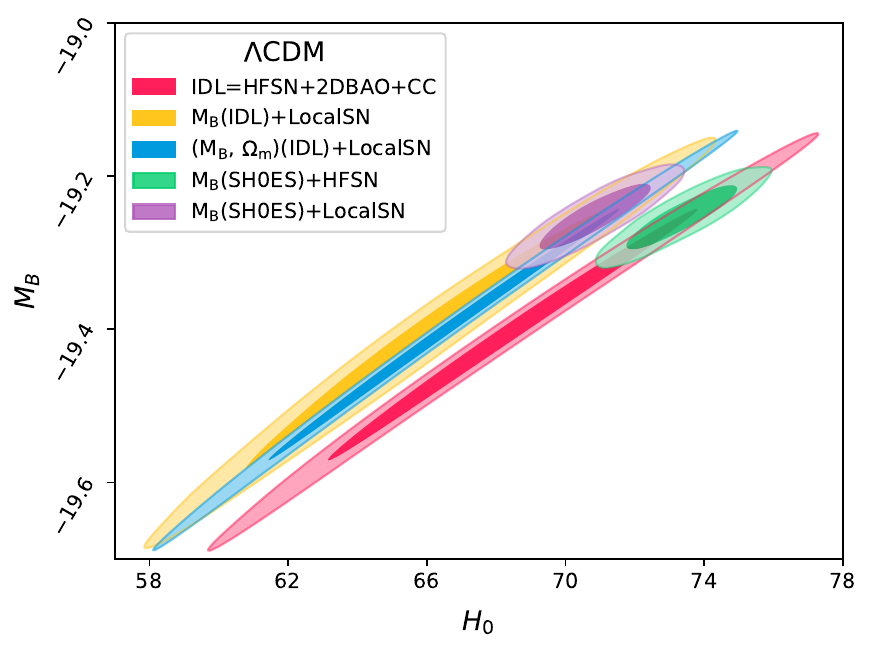}\\
\caption{Marginalized 1$\sigma$ and 2$\sigma$ contours of $H_0$ and $M_B$ from IDL-calibrated local SNe (yellow and blue) compared to the pure IDL constraint (red). The SH0ES-calibrated local SNe and HF SNe are shown in purple and green, respectively.}
\label{fig:LCDM-MB-H0}
\end{figure}

Taking the $\Lambda$CDM model as a typical example, we can infer the $H_0$ posterior from fitting local SNe calibrated by the IDL $M_B$ posterior with/without the IDL $\Omega_m$ posterior as an extra prior as shown in blue and yellow in Fig.~\ref{fig:LCDM-MB-H0}, respectively, where the pure IDL constraint is shown in red for comparison. It is then intriguing to observe that, although both $M_B$ and $H_0$ are consistent between IDL constraints and IDL-calibrated local constraints, they are actually separated sharply along the diagonal direction. This diagonal separation seems to be independent of calibrations as it also occurs even between the SH0ES-calibrated local SN and HF SN samples shown in purple and green, respectively.

\begin{figure*}
\centering
\includegraphics[width=0.32\textwidth]{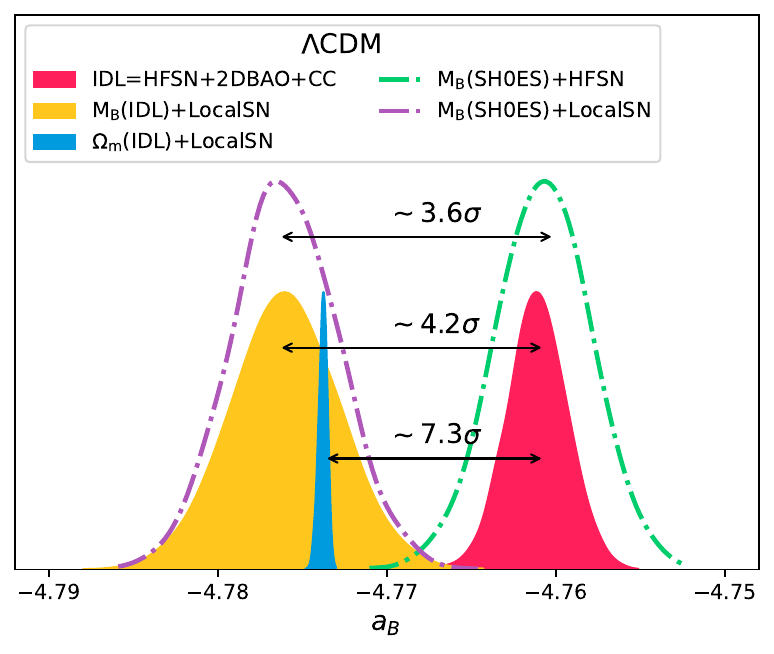}
\includegraphics[width=0.32\textwidth]{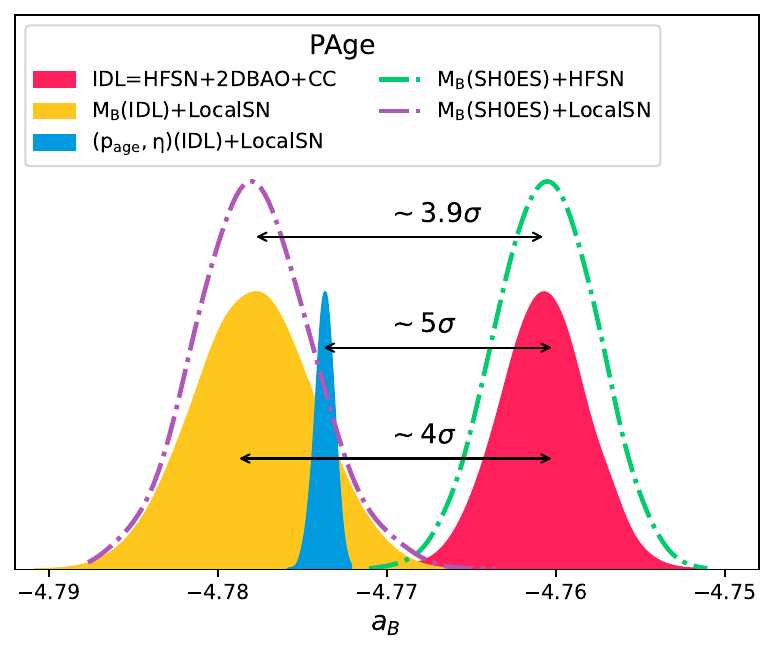}
\includegraphics[width=0.32\textwidth]{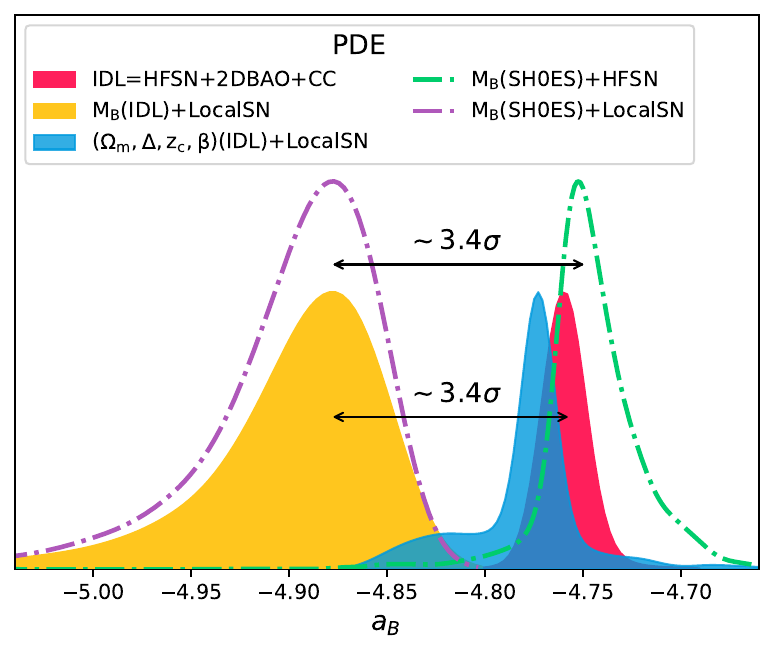}\\
\caption{1D posteriors of $a_B$ for $\Lambda$CDM, PAge, and PDE models from the pure IDL constraint (red), IDL-calibrated local direct (blue) and indirect (yellow) constraints, and SH0ES-calibrated constraints from HF SN (green) and local SN (purple).}
\label{fig:LCDM-PAge-aB}
\end{figure*}

This diagonal direction is in fact the intercept $a_B$ in the magnitude-redshift relation $m_{B,i}=5\lg d_L(z_i)-5a_B$ that is degenerated in both $H_0$ and $M_B$ by
\begin{align}\label{eq:aBdefinition}
a_B\equiv-\frac15\left(M_B+5\lg\frac{c}{H_0\cdot\mathrm{Mpc}}+25\right),
\end{align}
which can be directly inferred from $H_0$ and $M_B$ posteriors by fitting a model to IDL data sets. As a comparison, this $a_B$ inference can be computed by a $H_0$ posterior from fitting a given model to the local SN sample calibrated by the IDL posterior on $M_B$ as a prior. This $a_B$ inference can also be obtained from a local SN sample $m_{B,i}$ with a covariance matrix $C_{ij}$ by~\cite{Efstathiou:2021ocp}
\begin{align}\label{eq:aBcomputation}
a_B=\left(\sum_{ij}C_{ij}^{-1}\left(\lg d_L(z_i; \{p_i\})-\frac15m_{B,i}\right)\right)\bigg/\sum_{ij}C_{ij}^{-1},
\end{align}
which has been calibrated by IDL posteriors on model parameters $\{p_i\}=\{\Omega_m\}, \{p_\mathrm{age},\eta\}, \{\Omega_m,\Delta, z_c,\beta\}$ for the $\Lambda$CDM, PAge, and PDE models, respectively. The above full $a_B$ inferences are shown in Tab.~\ref{tab:FittingResultsaB}.
It is easy to see that this diagonal separation surprisingly leads to a dubbed \textit{$a_B$ tension} as also shown explicitly in Fig.~\ref{fig:LCDM-PAge-aB} for $\Lambda$CDM (left), PAge (middle), and PDE (right) models.

\begin{table}[H]
\caption{Inferences on $a_B$ from local and late Universe}
\label{tab:FittingResultsaB}
\renewcommand{\arraystretch}{1.5}
\centering
\begin{tabular}{|c|c|c|}
\hline
Samples           & Models & $a_B$ inference \\ \hline
\multirow{3}{*}{IDL=HFSN+2DBAO+CC} & $\Lambda$CDM   & $-4.7612\pm0.0018$            \\ \cline{2-3} 
                  & PAge    & $-4.7608\pm0.0026$\\ \cline{2-3} 
                  & PDE     & $-4.7614^{+0.0024}_{-0.0020}$           \\ \hline
\multirow{3}{*}{$M_B$(IDL)+Local SN} & $\Lambda$CDM & $-4.7761\pm0.0031$           \\ \cline{2-3} 
                  & PAge     & $-4.7780\pm0.0034$ \\ \cline{2-3} 
                  & PDE      & $-4.78^{+0.019}_{-0.061}$             \\ \hline
\multirow{3}{*}{$\{p_i\}$(IDL)+Local SN} & $\Lambda$CDM & $-4.7739^{+0.00027}_{-0.00023}$             \\ \cline{2-3} 
                  & PAge      &$-4.7738^{+0.00060}_{-0.00051}$                \\ \cline{2-3} 
                  & PDE      & $-4.905^{+0.058}_{-0.017}$             \\ \hline
\end{tabular}
\end{table}

The pure IDL constraints on $a_B$ are all shown in red via~\eqref{eq:aBdefinition} from IDL posteriors of $M_B$ and $H_0$, which are in $3\sim7\sigma$ tensions with the IDL-calibrated local constraints on $a_B$ for both $\Lambda$CDM and PAge models. Here, the IDL-calibrated local constraint on $a_B$ can be implemented either \textit{directly} by the IDL posterior(s) on the model parameter(s) via~\eqref{eq:aBcomputation} as shown in blue, or \textit{indirectly} by the IDL posterior on $M_B$ to first constrain $H_0$ and then $a_B$ via~\eqref{eq:aBdefinition} as shown in yellow. This $a_B$ tension occurs for both IDL-calibrated local direct and indirect constraints with respect to the pure IDL constraint except for the PDE case with more model parameters to enlarge the uncertainty in the IDL-calibrated local direct constraint. Nevertheless, a tension in $a_B$ between the IDL-calibrated local indirect constraint and the pure IDL constraint persists for the PDE model. Intriguingly, this $a_B$ tension seems to be independent of calibrations in use as it also occurs even between the SH0ES-calibrated local SN and HF SN samples shown in purple and green dashed-dotted curves, respectively.

Although the PAge model is comparable to the $\Lambda$CDM model in fitting the local Universe, we find decisive BIC evidence $\Delta\mathrm{BIC}=-36.9$ and decisive Bayesian factor $\Delta\ln\mathcal{Z}=16.4$ with \texttt{DYNESTY} supporting a PDE transition at $z_c=0.00218_{-0.00073}^{+0.00063}$ over the $\Lambda$CDM model. However, this PDE model still suffers from the $a_B$ tension as other models, and hence it seems suspicious to expect such an ultra-low redshift PDE transition in our nearby Universe. See, however, Refs.~\cite{Perivolaropoulos:2021bds,Perivolaropoulos:2022khd,Gomez-Valent:2023uof,Liu:2024vlt} for recent hints of $M_B$ evolution equivalent to a PDE transition but at a higher redshift. Besides, the Hubble diagram found in the Carnegie-Chicago Hubble Program VIII~\cite{Freedman:2019jwv} admits very little scatter in their velocity flow corrected distances of the SN host galaxies with redshift $z\lesssim0.007$, therefore, there seems no room for an abrupt change to the equation of state at very low redshifts~\cite{Efstathiou:2021ocp}. Therefore, this preference for a local PDE transition might as well be an illusion of some unknown local systematics.

\section{Looking into the $a_B$ tension}

Having obtained the $a_B$ tension regardless of models ($\Lambda$CDM, PAge, or PDE) and calibrations (direct/indirect calibrations from IDL posteriors, or SH0ES calibration), we now turn to source the causes of it. Recall the definition of the intercept $-5a_B=M_B+5\lg(c/H_0/\mathrm{Mpc})+25$ from the magnitude-redshift relation $m_{B,i}=5\lg d_L(z_i)-5a_B$, the very difficulty of the current debate on the Hubble tension is that any tension in $H_0$ can be attributed to either new physics in modeling $d_L(z)$ or systematics on calibrating $M_B$ in order to match the apparent magnitude $m_{B,i}$. Now that we have identified a tension in $a_B$ as a whole, then we only need to call for new physics in $d_L(z)$ so as to agree with observations of $m_{B,i}$ at redshifts $z_i$, which is independent of uncertainties in calibrating $M_B$. Therefore, the possible causes of the $a_B$ tension can only be traced back to the new physics in modeling $d_L(z)$ or the uncertainty in measuring $z_i$ that is contributed mainly from the peculiar velocity around $z\lesssim0.01$.

\begin{figure*}
\centering
\includegraphics[width=0.4\textwidth]{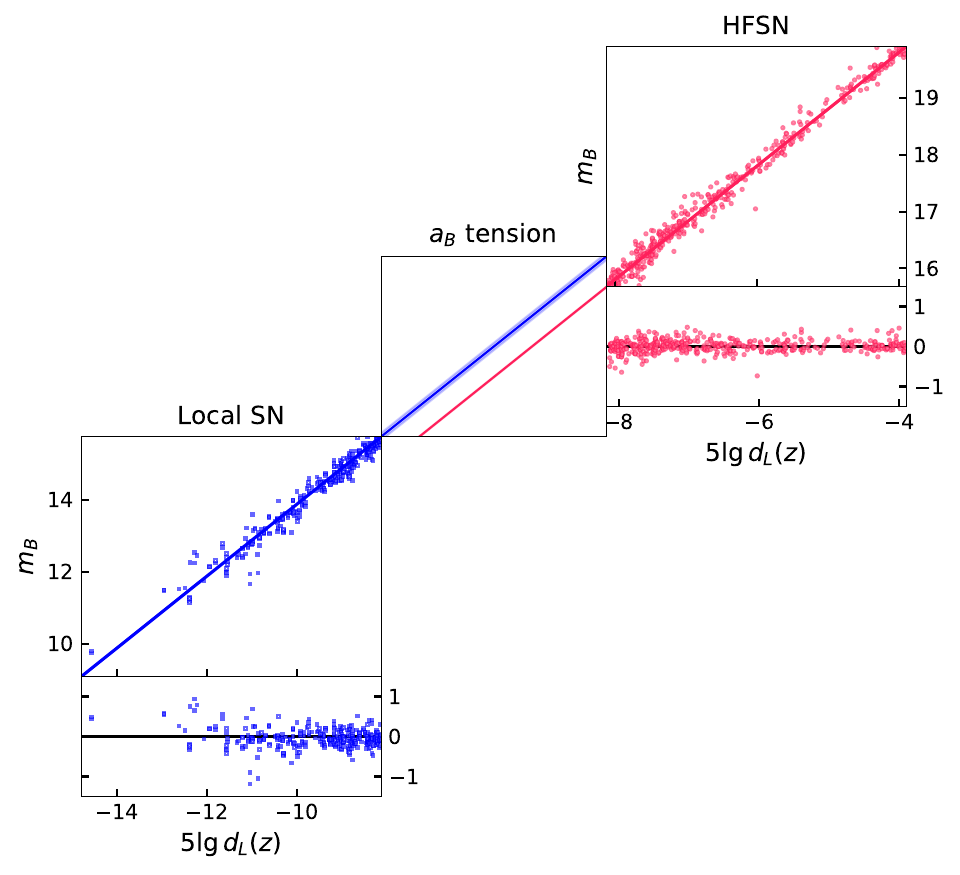}
\includegraphics[width=0.59\textwidth]{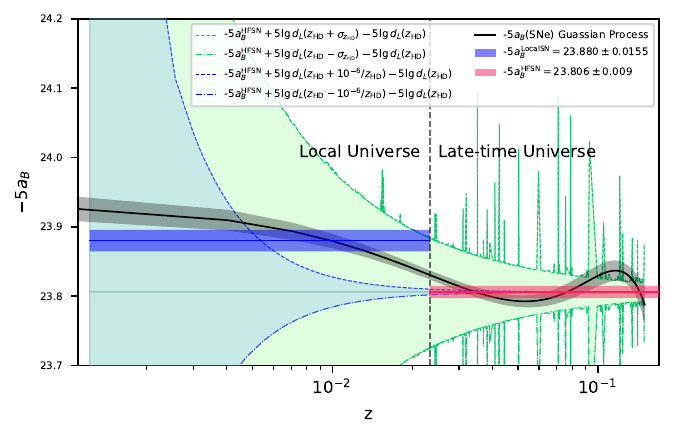}\\
\caption{The visual demonstration of the intercept tension between local and late Universe by constraining the intercepts $-5a_B$ from local and HF SN samples as shown in blue and red with $1\sigma$ errors for both panels, respectively. In the right panel, the volume effect of SN samples and the peculiar-velocity effect in the model are shown in light blue- and green-shaded regions with $1\sigma$ errors, respectively. The light black-shaded region is directly reconstructed from the SN data using the Gaussian process.}
\label{fig:dL-aB}
\end{figure*}

We then visualize these two possible origins of the $a_B$ tension in Fig.~\ref{fig:dL-aB}, where the blue- and red-shaded bands clearly depict the tension in $-5a_B$ with $1\sigma$ errors from constraining the local and HF SN samples, respectively. The light blue-shaded region accounts for the selection effect from the volumetric or Malmquist bias visualized effectively by the uncertainty in the model as $5\lg d_L(z_\mathrm{HD}+10^{-6}/z_\mathrm{HD})-5\lg d_L(z_\mathrm{HD})$~\cite{Kenworthy:2022jdh}, which is obviously insufficient to account for the $a_B$ tension. The peculiar velocity effect on the redshift measurement uncertainty is also not responsible for the $a_B$ tension as the estimated uncertainty $\sigma_{z_\mathrm{HD}}$ has already been incorporated into the total covariance matrix~\cite{Peterson:2021hel,Carr:2021lcj}. Nevertheless, suppose there is some new physics mimicking the effect of peculiar velocity, then its maximal impact on $-5a_B$ from the model of $d_L(z)$ at a redshift with $1\sigma$ errors reads $5\lg d_L(z_\mathrm{HD}\pm\sigma_{z_\mathrm{HD}})-5\lg d_L(z_\mathrm{HD})$, which is shown within a light green-shaded region that seems to fully cover the $a_B$ tension. Any new physics of resolutions to the $a_B$ tension should interpolate the blue and red bands within the light green-shaded region. As a typical example, we use the Gaussian process regression to model-independently reconstruct the intercept evolution directly from the Pantheon+ data with the help of an open source code \texttt{Scikit-learn}~\cite{scikit-learn,10.7551/mitpress/3206.001.0001,Seikel:2012uu,Gomez-Valent:2023uof}, which also exhibits a data-driven discrepancy in $a_B$ between the local and late Universe as shown in black with $1\sigma$ errors.


\section{Conclusions and discussions}

In this paper, we have further established the dubbed~\cite{Vagnozzi:2023nrq} late-time ``No-Go theorem''~\cite{Benevento:2020fev,Camarena:2021jlr,Efstathiou:2021ocp,Cai:2021weh,Cai:2022dkh} with a fully model-independent IDL constraint against any homogeneous new physics beyond the $\Lambda$CDM model in the late Universe. We then raise the $a_B$ tension between the local and late Universe as shown illuminatingly in Fig.~\ref{fig:dL-aB} independent of models and calibrations in use. This opens a new window of looking into the Hubble tension from a local perspective via, not $H_0$ or $M_B$ alone separately, but their combined $a_B$ as a whole to be a more sensitive observable and diagnostics. Future checks with other distance calibrators and indicators would be indispensable for confirmatively claiming the $a_B$ tension as a local reflection of the Hubble tension~\cite{Huang:2024gfw}.

Sourcing the origin of the $a_B$ tension simply leads to either new physics in modeling the dimensionless luminosity distance $d_L(z)$ or uncertainties in measuring the apparent magnitude $m_{B,i}$ at a redshift $z_i$ due to volumetric selection effect and peculiar velocity effect~\cite{Huang:2024gfw}. We have shown that the volume effect is irrelevant for the $a_B$ tension. This is different from a positive tail seen in the ultra-low redshift of the Hubble residuals of SNe~\cite{Brout:2022vxf} and Cepheid distances~\cite{Kenworthy:2022jdh}. Although the peculiar velocity effect has been incorporated into the covariance matrix, we can convert it into some model uncertainty, enclosing the possible range of new physics that might disguise itself as local systematics in the redshift uncertainty.

Local systematics do not need to come from our own local Universe but can also refer to a distant local Universe that impacts our local measurements. Such an impact should be necessarily small, but the observed correlation between the amount of such an impact and the property of that distant local Universe has recently been found to be in direct tension with the perturbative prediction from the $\Lambda$CDM model. This newly discovered \textit{$\delta H_0$ tension}~\cite{Yu:2022wvg} therefore signifies a cosmological breakdown of the $\Lambda$CDM model at perturbative level in the local-scale Universe, hinting for some local-scale new physics disguised as local systematics that would dim distance indicators and calibrators in denser regions~\cite{Cai:2021wgv,Giani:2023aor}. Yet its relation to the $a_B$ tension merits further studies.

\begin{acknowledgments}
We thank Sunny Vagnozzi for an early correspondence on the model dependency of 3D BAO analysis.
This work is supported by 
the National Key Research and Development Program of China Grants No. 2021YFA0718304, No. 2021YFC2203004, and No. 2020YFC2201501,
the National Natural Science Foundation of China Grants No. 12105344, No. 12235019, and No. 12047503, 
the Science Research Grants from the China Manned Space Project with No. CMS-CSST-2021-B01,
and the Postdoctoral Fellowship Program of CPSF.
We also acknowledge the use of the HPC Cluster of ITP-CAS.
\end{acknowledgments}

\bibliography{ref}

\end{document}